\documentclass[aps,11pt,preprintnumbers,nofootinbib,floatfix]{revtex4}
\usepackage{subfigure}
\usepackage{multirow}
\usepackage{amsmath}

\usepackage{amssymb}
\usepackage{graphicx}

\begin{document}

\title{Four-dimensional de Sitter cosmology on D-branes nucleated in an asymptotically $\text{AdS}_5\times T^{1,1}$ background}


\author{Cao H. Nam$^{1,2}$}
\email{caohoangnam@duytan.edu.vn}  
\affiliation{$^1$Institute of Theoretical and Applied Research, Duy Tan University, Hanoi 100000, Vietnam\\
$^2$School of Engineering and Technology, Duy Tan University, Da Nang 550000, Vietnam}
\date{\today}

\begin{abstract}
We find four-dimensional de Sitter (dS) vacuum solutions on probe D-branes nucleating in a
background, which is a U(1) charged black hole solution of IIB supergravity, including stringy corrections. A sufficiently high chemical potential induced by the wrapped D3-brane charge, inducing an instability in the system, is essential to lead to the nucleation of the probe D-branes. We show that stringy corrections can yield a dS vacuum on spherical D3-branes consistent with the observations without fine-tuning. Motivated by the fact that the matter fields propagating in the compact extra dimensions can provide solutions for problems in particle physics and cosmology, we also construct a dS vacuum on a probe D5-brane that wraps on a two-torus of the internal manifold $T^{1,1}$. This construction requires turning on a worldvolume U(1) gauge field along the wrapped part of the D5-brane and dissolving some D3-branes in the D5-brane. The stringy corrections play an important role in yielding a dS vacuum, and the field strength of the U(1) gauge field must be sufficiently large to produce a tiny cosmological constant.
\end{abstract}

\maketitle
\section{Introduction}
String theory has been recognized as a prominent candidate for an ultraviolet complete theory that may connect quantum gravity at the high-energy regime with particle physics and cosmology at the low-energy regime. One of the challenge questions is how a dimensional reduction of string theory can lead to a 4D effective field theory with a de Sitter (dS) vacuum. This is motivated by the measurements of the cosmic microwave background \cite{Aghanim2020} and the accelerating expansion of the universe at the late time \cite{Peebles2003,Lonappan2018}, indicating the existence of a positive vacuum energy density whose magnitude is about $(2.4\times10^{-3}\text{eV})^4$ \cite{Tanabashi2018}. Hence, many attempts have been made to construct dS vacua in the string theory compactification. The (metastable) dS vacua can be found based on the uplift of anti-de Sitter (AdS) vacua via the anti D$p$-branes \cite{Kachru2003,Balasubramanian2005}, the non-trivial behavior of 4D graviton in the extra dimensions \cite{Nam2023}, and the decay of the non-supersymmetric AdS vacuum \cite{Banerjee2018}. So far, constructing a dS vacuum that is at least metastable and consistent with the observations is still an outstanding issue in string theory \cite{Danielsson2018}.

The D$p$-branes on which open strings can end with the Dirichlet boundary condition along the transverse directions have been used to realize the standard model (SM) as well as physics beyond the SM \cite{Aldazabal2001,Blumenhagen2005,Anchordoqui2012,Antoniadis2003}. They are the BPS states that forbid corrections to their tension, which is the energy per volume of the brane. Therefore, cosmology on a BPS D$p$-brane (which is wrapped on a $(p-3)$ cycle in the internal space with $p>3$) describes an expanding universe with a zero vacuum energy density. This provides a mechanism to guarantee that the vacuum energy density on the D$p$-brane worldvolume does not obtain large quantum corrections. However, when the D$p$-brane moves in the higher-dimensional bulk space with supersymmetry broken, there are non-zero stringy corrections that originate from the finite size of strings. It is expected that these corrections would lead to a non-vanishing but small vacuum energy density because the stringy corrections are highly suppressed by the string length $\sqrt{\alpha'}$.

The near-horizon limit of the geometry of 10D bulk spacetime curved by a stack of D3-branes is a direct product $\text{AdS}_5\times E_5$ where $E_5$ is a 5D Sasaki-Einstein compact manifold. The well-known case is that the manifold $E_5$ is a five-sphere $S^5$, corresponding to that the maximal number of supersymmetries is preserved. In the limit that the number of D3-branes is very large \cite{Hooft1974}, the gravity propagating in the $\text{AdS}_5\times S^5$ background is dual to the $\mathcal{N}=4$ gauge theory on the stack of D3-branes \cite{Maldacena1997}. Although the $\text{AdS}_5\times S^5$ background is very simple for the holographic investigations, it is less interesting to build cosmological and particle physics models because completely breaking the maximal number of supersymmetries is, in general, hard to achieve. In other words, the 10D stringy background should have fewer unbroken supersymmetries for the phenomenological interest. If the internal space $E_5$ is the $T^{1,1}$ manifold \cite{Romans1985}, then the dual gauge theory on the stack of D3-branes has the minimal $\mathcal{N}=1$ supersymmetry \cite{Klebanov1998}. 

Taking into account a finite density corresponding to a non-zero chemical potential, the authors in \cite{Herzog2010} extended the $\text{AdS}_5\times T^{1,1}$ geometry by turning on a baryonic U(1) gauge field which is sourced by D3-branes wrapping on topologically non-trivial $3$-cycles of $T^{1,1}$. The baryonic U(1) symmetry leads to a chemical potential difference in the bosons and fermions, by which it provides a way to break $\mathcal{N}=1$ supersymmetry. It is interesting that when the chemical potential is large enough, a probe D3-brane hidden by the horizon of the charged black hole can nucleate and expand along the radial direction of the AdS$_5$ part of the 10D bulk. This can be realized by studying the effective potential of a probe D3-brane \cite{Henriksson2020}. This implies the D3-brane emission from the stack of D3-branes in analogy with the Hawking radiation in conventional black holes.

In this paper, we identify our observable universe living on some D-branes that nucleate in the unstable background mentioned above. In Sec. \ref{D3-nucle}, we compute the effective action for the D3-brane and then obtain the Friedmann equation describing its expansion at the late time at the leading order of string length. We show that at this order, no self-gravity accelerates the expansion of the D3-brane corresponding to the vanishing cosmological constant on its worldvolume. In Sec. \ref{alp-corr}, we consider stringy corrections up to the order $\alpha'^2$, leading to a positive cosmological constant $\sim$ (string length/AdS radius)$^4$. In this way, if the string length is small enough and the AdS radius is large enough, the stringy corrections can lead to a tiny cosmological constant as observed by the experiments. However, there is a relation between the observed 4D Planck scale and the parameters of the string length and the AdS radius. Hence, the ratio of the string length to the AdS radius is small enough, leading to the observed value of the 4D cosmological constant, only in the case that the spatial curvature of the D3-brane worldvolume is positive. 

Considering the SM fields and the fields beyond the SM that can propagate in the extra dimensions can provide solutions for puzzles in cosmology and particle physics \cite{Ponton2011}. Hence, in Sec. \ref{D5-nucle}, we explore the nucleation of a probe D5-brane which wraps on a compact two-dimensional manifold of $T^{1,1}$ with a two-torus as an example without loss of generality. The nucleation of the D5-brane in the background that has only the non-zero four-form potential requires turning on the field strength two-form on its worldvolume and some D3-branes dissolving in the D5-brane. We show that the stringy corrections are essential to yield a 4D dS vacuum, and the field strength two-form must be large enough to produce a tiny cosmological constant.

\section{\label{D3-nucle}Nucleated D3-brane}

The near-horizon limit of the stack of D3-branes placed at a conifold tip (given by $\sum^4_{i=1}z^2_i=0$ with $z_i\in\mathbb{C}$ whose base is $T^{1,1}$) leads to the 10D geometry of the form $\text{AdS}_5\times T^{1,1}$ \cite{Klebanov1998}. We add the D3-branes, which wrap entirely on a topologically non-trivial $3$-cycle of the internal manifold $T^{1,1}$, by which they are stable and behave as charged point particles from the (external) five-dimensional viewpoint. These wrapped D3-branes are the source of the four-form gauge field $C_4$, which has three indices along the $3$-cycle and one index along the time direction $\mathcal{T}$, corresponding to a U(1) gauge field. This brane configuration allows us to find a U(1) charged black hole solution of IIB supergravity given by \cite{Herzog2010}
\begin{eqnarray}
ds^2_{10}&=&e^{-\frac{5}{3}\chi(R)}ds^2_5+L^2e^{\chi(R)}\left[\frac{e^{\eta(R)}}{6}\sum^2_{i=1}\left(d\Theta^2_i+\sin^2\Theta_id\Phi^2_i\right)+\frac{e^{-4\eta(R)}}{9}g^2_5\right],\label{10D-back-metri}\\
ds^2_5&=&-g(R)e^{-w(R)}d\mathcal{T}^2+\frac{dR^2}{g(R)}+R^2d\Omega^2_{3,k},\label{5D-back-metri}
\end{eqnarray}
where $g_5=d\Psi+\cos\Theta_1d\Phi_1+\cos\Theta_2d\Phi_2$, $L$ is the radius of the asymptotically AdS space given by $ds^2_5$ and is related to the number of the D3-branes $N$ placed at the conifold tip as $L^4=4\pi g_sN(\alpha')^2 (27/16)$, and 
$d\Omega^2_{3,k}$ is the line element of the horizon with the topology determined by the curvature $k$. 

The four-form gauge field $C_4$ corresponding to the geometry given by (\ref{10D-back-metri}) is
\begin{equation}
C_4=\left(C_4\right)_\mathcal{T} d\mathcal{T}\wedge \text{vol}_3+\frac{L^3}{9\sqrt{2}}A_\mathcal{T}d\mathcal{T}\wedge\omega_2\wedge g_5,\label{C4-from-pot}   
\end{equation}
where $\text{vol}_3$ denotes the volume form corresponding to the line element $d\Omega^2_{3,k}$, the components $\left(C_4\right)_\mathcal{T}$ and $A_\mathcal{T}$ are determined by
\begin{eqnarray}
\left(C_4\right)_\mathcal{T}&=&\frac{4}{L}\int^R_{R_h}r^3e^{-\frac{w(r)}{2}-\frac{20\chi(r)}{3}}dr,\\
A_\mathcal{T}&=&\int^R_{R_h}\frac{Q}{r^3}e^{-\frac{w(r)}{2}-2\eta(r)+\frac{4}{3}\chi(r)}dr,
\end{eqnarray}
with $R_h$ and $Q$ being the black hole horizon determined by $g(R_h)=0$ and the black hole charge, respectively, and the two-form 
$\omega_2$ is given by
\begin{eqnarray}
\omega_2=\frac{1}{2}\left(\sin\Theta_1 d\Theta_1\wedge d\Phi_1-\sin\Theta_2 d\Theta_2\wedge d\Phi_2\right).   
\end{eqnarray}
The self-dual five-form $F_5$ corresponding to the four-form gauge field $C_4$ is determined by $F_5=\left(\mathcal{F}+\star\mathcal{F}\right)/g_s$ where $\star$ denotes the Hodge dual and $\mathcal{F}=dC_4$ is obtained as follows
 \begin{eqnarray}
 \mathcal{F}=\frac{4}{L}e^{-\frac{20}{3}\chi(R)}\text{vol}_5+\frac{L^3}{9\sqrt{2}}\frac{dA_\mathcal{T}}{dR}dR\wedge d\mathcal{T}\wedge\omega_2\wedge g_5,\label{5-form-dC4}
 \end{eqnarray}
where $\text{vol}_5$ denotes the volume form associated with the 5D line element $ds^2_5$. Note that other $(p+1)$-form fluxes of IIB supergravity are zero.

The field strength tensor $F=A'_\mathcal{T}(R)d\mathcal{T}$ of the U(1) gauge field produces a warped product in the 10D bulk geometry, breaks the isometry of AdS$_5$, and squashes the internal manifold $T^{1,1}$. These are governed by the functions $\chi(R)$, $w(R)$, and $\eta(R)$, respectively. These functions and the function $g(R)$ are given in the expansion form in the region of $L/R\ll1$ by \cite{Herzog2010}
\begin{eqnarray}
\chi(R)&=&-\frac{Q^2L^2}{200R^6}+\frac{\chi_8L^8}{R^8}+\cdots,\nonumber\\
w(R)&=&\mathcal{O}\left(\frac{L^{12}}{R^{12}}\log^2\left(\frac{R}{L}\right)\right),\nonumber\\
\eta(R)&=&\frac{Q^2L^2}{80R^6}\log\left(\frac{R}{L}\right)+\frac{\eta_6L^6}{R^6}+\cdots,\label{asym-behav}\\
g(R)&=&k+\frac{R^2}{L^2}+\frac{g_2L^2}{R^2}+\frac{Q^2L^4}{12R^4}+\cdots\nonumber,
\end{eqnarray}
where the ellipses refer to the higher-order terms in $L/R$ and the constants $\chi_8$, $\eta_6$, and $g_2$ are the expansion factors.

In this work, we consider a physical system that is determined by Eqs. (\ref{10D-back-metri}), (\ref{5D-back-metri}) and (\ref{C4-from-pot}). A particular point of this system is that it can exhibit instability, which is indicated by computing the effective potential for a probe D3-brane, given in \cite{Henriksson2020}. It was found that the instability would appear when the chemical potential $\mu=A_\mathcal{T}(\infty)-A_\mathcal{T}(R_h)$ corresponding to the U(1) gauge field $A$ is large enough and the black hole temperature $T$ is sufficiently low. In this situation, there is a potential barrier between the regions inside and outside the black hole horizon. As a result, a probe D3-brane inside the horizon would tunnel across the potential barrier towards a global minimum located at $R\rightarrow\infty$. This implies that the physical system would emit some D3-branes from the stack of D3-branes placed at the conifold tip to reach the stable state. This leads to the nucleation of some D3-branes, which expand in the radial direction of the asymptotically AdS$_5$ space. The ratio of the black hole temperature $T$ to the chemical potential $\mu$ is essential to the nucleation of the D3-brane. It was indicated that the nucleation rate per volume of the D3-brane, $\Gamma/V$, is different from zero for $T/\mu\lesssim0.2$, but it would approach zero when $T/\mu\approx0.2$ \cite{Henriksson2022}. 

We identify the nucleated D3-brane as the observable universe where $d\Omega^2_{3,k}$ given in Eq. (\ref{5D-back-metri}) determines the spatial curvature of the brane universe. To be consistent with the observations of the spatial curvature \cite{Planck2020}, we consider the horizon which has the spherical topology ($k=1$) and the planar topology ($k=0$). These horizon topologies correspond to $d\Omega^2_{3,1}=\left[d\psi^2+\sin^2\psi\left(d\theta^2+\sin^2\theta d\phi^2\right)\right]$ and $d\Omega^2_{3,0}=dx^2+dy^2+dz^2$. Note that in Refs. \cite{Herzog2010,Henriksson2020,Henriksson2022}, the horizon geometry of the black hole, which was considered, is flat; however, the analysis is valid for the general geometry of the horizon because the horizon curvatures of $k=1$ and $k=0$ correspond to the global AdS and Poincare coordinate systems of the AdS space, respectively.

The embedding of the D3-brane into the background given by Eqs. (\ref{10D-back-metri}), (\ref{5D-back-metri}) and (\ref{C4-from-pot}) is described by the following coordinate functions  
\begin{eqnarray}
\mathcal{T}=\mathcal{T}(t), \ \  R=R(t), \ \ \Vec{x}\Big|_{\text{bulk}}=\Vec{x}\Big|_{\text{brane}},\ \ X^i\Big|_{T^{1,1}}=\text{constant},\label{D3-br-embe}
\end{eqnarray}
where $t$ is the time coordinate on the D3-brane, $\Vec{x}$ denotes the coordinates $(\psi,\theta,\phi)$ for $k=1$ and $(x,y,z)$ for $k=0$, and $X^i\Big|_{T^{1,1}}$ refer to the coordinates on the manifold $T^{1,1}$. Then the induced metric on the D3-brane reads
\begin{eqnarray}
ds^2_4&=&G_{MN}e^M_\mu e^N_\nu dx^\mu dx^\nu\nonumber\\
&=&e^{-5\chi(a)/3}\left(-\gamma_{tt}dt^2+a^2d\Omega^2_{3,k}\right),\label{D3-br-metri}
\end{eqnarray}
where $e^M_\mu=\partial X^M/\partial x^\mu$ are the tangent vectors to the D3-brane worldvolume, $a$ is the scale factor related to the radial coordinate $R$ as $a(t)\equiv R(t)$, and $\gamma_{tt}$ is given by
\begin{eqnarray}
\gamma_{tt}=g(a)e^{-w(a)}\dot{\mathcal{T}}^2-\frac{\dot{a}^2}{g(a)},    
\end{eqnarray} 
which approaches the unit at the late time.

We study the expansion of the D3-brane at the late time corresponding to the region of $L/a\ll1$. First, let us compute the D3-brane action which is given as
\begin{equation}
 S_{\text{D}3}=-T_3\int d^4x\sqrt{|g|}+T_3\int P[C_4],  
\end{equation}
where $g\equiv\det(g_{\mu\nu})$, $T_3=(2\pi)^{-3}/(g_s\alpha'^2)$ is the tension of the D3-brane, and $P[C_4]$ refers to the pullback of the four-form potential $C_4$ to the D3-brane worldvolume. Note that the dilaton term $e^{-\Phi}=1/g_s$ has been absorbed into the brane tension. At the late time ($L/a\ll1$) and with $L/R_h\ll1$, we can use the asymptotic behavior of the functions $\chi(a)$ and $w(a)$ given in Eq. (\ref{asym-behav}) to compute the pullback of the four-form potential $C_4$, which reads
\begin{eqnarray}
P[C_4]&=&\left(\frac{4}{L}\int^{a}_{R_h}r^3e^{-\frac{w(r)}{2}-\frac{20\chi(r)}{3}}dr\right)\dot{\mathcal{T}}dt\wedge dx\wedge dy\wedge dz\nonumber\\
&=&\frac{1}{L}\left[a^4-\frac{Q^2L^2}{15a^2}-\left(R^4_h-\frac{Q^2L^2}{15R^2_h}\right)+\cdots\right]\dot{\mathcal{T}}dt\wedge dx\wedge dy\wedge dz,
\end{eqnarray}
where the ellipsis refers to the higher-order terms in $L/a$. Then, the D3-brane action at the late time is found as 
\begin{eqnarray}
S_{\text{D}3}&=&-T_3V_3\int dt\left\{\sqrt{\gamma_{tt}}e^{-\frac{10}{3}\chi(a)}a^3-\frac{1}{L}\left[a^4-\frac{Q^2L^2}{15a^2}-\left(R^4_h-\frac{Q^2L^2}{15R^2_h}\right)+\cdots\right]\dot{\mathcal{T}}\right\}\nonumber\\
&\equiv&\int dt\mathcal{L}_{\text{D}3},\label{tre-D3-act}
\end{eqnarray}
where $V_3\equiv\int d\Omega_{3,k}$ and $\mathcal{L}_{\text{D}3}$ is the Lagrangian of the D3-brane.

From the Euler-Lagrange equation for the D3-brane and $\partial\mathcal{L}_{\text{D}3}/\partial\mathcal{T}=0$, we find that the conjugate momentum $\partial\mathcal{L}_{\text{D}3}/\partial\dot{\mathcal{T}}$ (which is nothing but the energy) is conserved. It is equal to zero for an observer who observes the nucleation of the D3-brane at rest, corresponding to the following equation
\begin{equation}
\frac{e^{-\frac{10}{3}\chi(a)-w(a)}a^3g(a)\dot{\mathcal{T}}}{\sqrt{\gamma_{tt}}}-\frac{1}{L}\left[a^4-\frac{Q^2L^2}{15a^2}-\left(R^4_h-\frac{Q^2L^2}{15R^2_h}\right)+\cdots\right]=0.    
\end{equation}
With respect to the observer moving together with the nucleated D3-brane, corresponding to the proper time $\tau$ satisfying $d\tau=\sqrt{\gamma_{tt}}dt$, we have
\begin{eqnarray}
\dot{\mathcal{T}}=\sqrt{\frac{e^{w(a)}}{g(a)}\gamma_{tt}\left[1+\frac{(\partial_\tau a)^2}{g(a)}\right]}.
\end{eqnarray}
Finally, we determine the Friedmann equation governing the expansion of the D3-brane at the late time as
\begin{equation}
\left(\frac{\dot{a}}{a}\right)^2+\frac{k}{a^2}=\frac{L^2}{2}\left(\frac{Q^2}{15L^2R^2_h}-\frac{R^4_h}{L^4}-2g_2\right)\frac{1}{a^4}+\mathcal{O}\left(\frac{1}{a^6}\right),\label{Fr-equa}
\end{equation}
where the dot denotes the derivative with respect to the proper time $\tau$. The equation (\ref{Fr-equa}) describes the expansion of the probe D3-brane with the spatial curvature $k$ at the late time without having a positive cosmological constant.

\section{\label{alp-corr} $\alpha'^2$-corrections to D3-brane}

We study the nucleation of the D3-brane, including the corrections at the order $\alpha'^2$ for the brane action. At this order, the DBI action of the D-brane can obtain the curvature corrections \cite{Bachas1999}, whereas the higher-order derivatives of the fluxes can lead to the corrections to the WZ action \cite{Cheung1998,Jalali2016}. The D3-brane action, including these corrections, thus reads
\begin{eqnarray}
S^{\text{tot.}}_{\text{D}3}&=&S_{\text{D}3}+S^{(\alpha'^2)}_{\text{D}3}=-T_3\int d^4x\sqrt{|g|}\left\{1-\frac{\pi^2\alpha'^2}{48}\left[\left(R_T\right)_{\mu\nu\rho\lambda}\left(R_T\right)^{\mu\nu\rho\lambda}-2\left(R_T\right)_{\mu\nu}\left(R_T\right)^{\mu\nu}\right.\right.\nonumber\\
&&\left.\left.-\left(R_N\right)_{\mu\nu ab}\left(R_N\right)^{\mu\nu ab}+2\bar{R}_{ab}\bar{R}^{ab}\right]\right\}+T_3\left[\int P[C_4]\wedge\left(1+\frac{\pi^2\alpha'^2}{24}\left(\text{tr}\mathcal{R}^2_T-\text{tr}\mathcal{R}^2_N\right)\right)\right.\nonumber\\
&&\left.-\frac{\pi^2\alpha'^2}{12}\int d^4x\frac{\epsilon^{\mu\nu\rho\lambda}}{4!}\nabla_b\left(F_5\right)_{a\mu\nu\rho\lambda}\bar{R}^{ab}\right],\label{corr-D3-brane-act}
\end{eqnarray}
where the letters $a,b,...$ are the indices corresponding to the normal bundle, $\mathcal{R}_T$ ($\mathcal{R}_N$) is the curvature two-form of the tangent (normal) bundle, and $\epsilon^{\mu\nu\rho\lambda}$ is the Levi-Civita tensor on the D3-brane worldvolume. Note that because the curvatures squared $\mathcal{R}^2_T$ and $\mathcal{R}^2_N$ are the four-forms, the term $\sim\left(\text{tr}\mathcal{R}^2_T-\text{tr}\mathcal{R}^2_N\right)$ does not contribute to the correction of the WZ action, unless the zero-form potential $C_0$ is non-zero.

The quantities $(R_T)_{\mu\nu\rho\lambda}$ and $(R_N)_{\mu\nu ab}$ are the Riemann tensor on the D3-brane worldvolume and the projected curvature on the normal bundle, respectively. They are generally determined by both the pull-backs of the 10D Riemann tensor $R_{MNPQ}$ and the second fundamental form $K^a_{\mu\nu}$ via the Gauss-Codazzi equations as follows
\begin{eqnarray}
(R_T)_{\mu\nu\rho\lambda}&=&R_{MNPQ}e^M_\mu e^N_\nu e^P_\rho e^Q_\lambda+\delta_{ab}\left(K^a_{\mu\rho}K^b_{\nu\lambda}-K^a_{\mu\lambda}K^b_{\nu\rho}\right),\\
{\left(R_N\right)_{\mu\nu}}^{ab}&=&-R_{MNPQ}E^{M,a}E^{N,b}e^P_\mu e^Q_\nu+g^{\rho\lambda}\left(K^a_{\mu\rho}K^b_{\nu\lambda}-K^b_{\mu\rho}K^a_{\nu\lambda}\right),
\end{eqnarray}
where $E^M_a$ are the vectors forming an orthonormal frame for the normal bundle, and hence they satisfy the following relations
\begin{eqnarray}
E^M_aE^N_bG_{MN}=\delta_{ab}\ \ \ \ \ \ \text{and} \ \ \ \ \ \ E^M_ae^N_\mu G_{MN}=0.    
\end{eqnarray}
(Note that both $e^M_\mu$ and $E^M_a$ allow us to decompose the 10D bulk metric $G^{MN}$ as $G^{MN}=e^M_\mu e^N_\nu g^{\mu\nu}+E^M_aE^N_b\delta^{ab}$. The indices $a,b,...$ are raised and lowered by the metric $\delta_{ab}$ and its inverse $\delta^{ab}$ of the normal bundle.) The second fundamental form $K^a_{\mu\nu}$ can be written as
\begin{eqnarray}
K^a_{\mu\nu}&=&\left[\partial_\mu e^M_\nu-\left(\Gamma_T\right)^\rho_{\mu\nu}e^M_\rho+\Gamma^M_{PQ}e^P_\mu e^Q_\nu\right]E^{N,a}G_{MN}\equiv\Omega^M_{\mu\nu}E^{N,a}G_{MN},    
\end{eqnarray}
where $\left(\Gamma_T\right)^\rho_{\mu\nu}$ is the Christoffel symbols on the D3-brane worldvolume and is determined by the induced metric $g_{\mu\nu}$. 

The curvature tensor $\bar{R}_{ab}$ is given by
\begin{eqnarray}
\bar{R}_{ab}&=&g^{\mu\nu}R_{MPNQ}e^M_\mu e^N_\nu E^P_aE^Q_b+g^{\mu\nu}g^{\rho\lambda}\delta_{ac}\delta_{bd}\left(K^c_{\mu\rho}K^d_{\nu\lambda}-K^c_{\mu\nu}K^d_{\rho\lambda}\right).
\end{eqnarray}
The term $\sim K^c_{\mu\rho}K^d_{\nu\lambda}$ can be found from the disk-level scattering amplitudes \cite{Bachas1999} and the terms $\sim K^c_{\mu\nu}K^d_{\rho\lambda}$ can be determined by the $T$ and $S$ dualities \cite{Jalali2015}.

With the induced metric on the D3-brane worlvolume given by Eq. (\ref{D3-br-metri}), one can compute the terms related to the curvatures on the D3-brane worlvolume as follows
\begin{eqnarray}
\left(R_T\right)_{\mu\nu\rho\lambda}\left(R_T\right)^{\mu\nu\rho\lambda}-2\left(R_T\right)_{\mu\nu}\left(R_T\right)^{\mu\nu}&=&-12e^{\frac{10}{3}\chi(a)}\left\{\frac{k}{a^2}+\left(\frac{\dot{a}}{a}\right)^2\left[1-\frac{5}{2}a\chi'(a)+\frac{25}{36}a^2\chi'(a)^2\right.\right.\nonumber\\
&&\left.\left.-\frac{5}{6}a^2\chi''(a)\right]+\frac{\ddot{a}}{a}\left[1-\frac{5}{6}a\chi'(a)\right]\right\}^2.
\end{eqnarray}
The quantities $\dot{a}/a$, $1/a$, and $\ddot{a}/a$ at the present time are related to the Hubble constant $H_0$, the curvature density $\Omega_{k,0}$, and the cosmological constant $\Lambda_4$ as
\begin{eqnarray}
\frac{\dot{a}}{a}&=&H_0,\nonumber\\
\frac{1}{a}&=&H_0\sqrt{-\Omega_{k,0}},\\
\frac{\ddot{a}}{a}&=&\frac{\Lambda_4}{3}\sim H^2_0,\nonumber
\end{eqnarray}
where their experimental values are $H_0\approx2.5\times10^{-5}$ Mpc$^{-1}$ and $\Omega_{k,0}\approx-0.0438$ \cite{Planck2020}. As seen below, the experimental constraints impose an upper bound for the AdS radius $L$ of order $10^{-4}$ m. This suggests that the terms related to the curvatures of the D3-brane worldvolume at the late time are negligible compared to the terms $\sim1/L^4$.

The computation of the scalar curvatures of $\left(R_N\right)_{\mu\nu ab}\left(R_N\right)^{\mu\nu ab}$ and $\bar{R}_{ab}\bar{R}^{ab}$ is given in Appendix A, and they are expanded in terms of $L/a$ as follows
\begin{eqnarray}
\left(R_N\right)_{\mu\nu ab}\left(R_N\right)^{\mu\nu ab}&=&0,\\
\bar{R}_{ab}\bar{R}^{ab}&=&\frac{256}{L^4}\left[1+\frac{3k}{4}\left(\frac{L}{a}\right)^2+\frac{9k^2}{64}\left(\frac{L}{a}\right)^4-\frac{\left(2Q^2+47\right)}{96}\left(\frac{L}{a}\right)^6+\cdots\right].\label{RabRab} 
\end{eqnarray}
As we see below, the first term $\sim1/L^4$ in the right-hand side of Eq. (\ref{RabRab}), which is independent of $L/a$, would contribute to the cosmological constant on the D3-brane worldvolume. Whereas other terms, depending on $L/a$, lead to corrections to the spatial curvature, the energy density of the radiation, and the exotic components.
In addition, the correction for the WZ action is zero due to the equation of motion for $F_5$ as $d*F_5=0$.

The above results yield the nucleated D3-brane action up to the $\alpha'^2$-corrections at the late time as
\begin{eqnarray}
S^{\text{tot.}}_{\text{D}3}&\simeq&-T_3V_3\int dt\left\{\sqrt{\gamma_{tt}}e^{-\frac{10}{3}\chi(a)}a^3\left(1-\frac{32\pi^2\alpha'^2}{3L^4}\right)-\frac{1}{L}\left[a^4-\frac{Q^2L^2}{15a^2}-\left(R^4_h-\frac{Q^2L^2}{15R^2_h}\right)\right]\right\}.
\end{eqnarray}
This Lagrangian indicates that the brane tension is shifted as
\begin{eqnarray}
T_3\longrightarrow T_3\left(1-\frac{32\pi^2\alpha'^2}{3L^4}\right). 
\end{eqnarray}
This means that the decay of the unstable state with a sufficiently high chemical potential is mediated by the nucleation of the brane whose tension is smaller than its charge. This is compatible with the Weak Gravity Conjecture that is applied for a $p$-brane \cite{Palti2019}. 

The equation of motion for the D3-brane leads to
\begin{eqnarray}
\left(\frac{\dot{a}}{a}\right)^2+\frac{k}{a^2}\simeq\frac{64\pi^2\alpha'^2}{3L^6}+\frac{L^2}{2}\left[\left(\frac{Q^2}{15L^2R^2_h}-\frac{R^4_h}{L^4}\right)-2g_2\right]\frac{1}{a^4}.\label{corr-bubb-equa} 
\end{eqnarray}
This implies that the brane expansion at the late time is governed by a positive cosmological constant $\Lambda_4\sim\alpha'^2/L^6$ corresponding to a 4D dS vacuum. From Eq. (\ref{corr-bubb-equa}), we can extract a 4D vacuum energy density $\rho_4$ as
\begin{eqnarray}
\rho_4=\frac{64\pi^2\alpha'^2}{L^6}M^2_{\text{Pl}},
\end{eqnarray}
where $M_{\text{Pl}}$ is the 4D Planck scale. For $k=0$, $M_{\text{Pl}}$ is determined by using the projection of the bulk gravity on the brane \cite{Clifton2012}
\begin{eqnarray}
 M^2_{\text{Pl}}&\simeq&M^3_5L\nonumber\\
 &=&\frac{2L^6}{(2\pi)^7g^2_s\alpha'^4}\text{Vol}(T^{1,1}),
\end{eqnarray}
where $M_5$ is the 5D Planck scale of 10D gravity reduced on the internal manifold $T^{1,1}$ and $\text{Vol}(T^{1,1})=16\pi^3/27$ is the volume of $T^{1,1}$. 

For $k=1$, we can determine $M_{\text{Pl}}$ by identifying the spherical D3-brane as a nucleated bubble mediating the decay of a false AdS$_5$ vacuum into a true AdS$_5$ vacuum in such a way that \cite{Danielsson2023,Panizo2024}
\begin{eqnarray}
nT_3=\frac{3M^3_5(L_+-L_-)}{L_-L_+},\label{zero-4DCC}   
\end{eqnarray}
where $L_-$ and $L_+$ that approximate $L$ are the AdS radii inside and outside the bubble, respectively, and $n$ is the number of D3-branes emitted from the stack of $N$ D-branes. Note that the true and false AdS$_5$ vacua are inside and outside the bubble, respectively. The equation (\ref{zero-4DCC}) corresponds to the fact that the cosmological constant on the bubble vanishes at the leading order of $\alpha'$. This leads to $L_+-L_-\simeq2nL/(3N)$. By using the second Israel junction condition $\Delta K^{(-)}_{\mu\nu}-\Delta K^{(+)}_{\mu\nu}=T_{\mu\nu}/M^3_5$ where $\Delta K^{(\pm)}_{\mu\nu}=K^{(\pm)}_{\mu\nu}-g_{\mu\nu}K^{(\pm)}$ is related to the extrinsic curvatures and  $T_{\mu\nu}$ is the energy-momentum tensor on the bubble \cite{Nam2024}, we find $M_{\text{Pl}}$ in terms of the number of D3-branes placed at the conifold singularity, the number of D3-branes escaping from the stack, and the AdS radius $L$ as follows
\begin{eqnarray}
M^2_{\text{Pl}}&=&\frac{M^3_5}{2}\left(L_+-L_-\right)\nonumber\\
&\simeq&\frac{nL^6}{324\pi^4g^2_sN\alpha'^4}=\frac{9nN}{64\pi^2 L^2}.
\end{eqnarray}

\begin{figure}[ht]
  \includegraphics[scale=0.4]{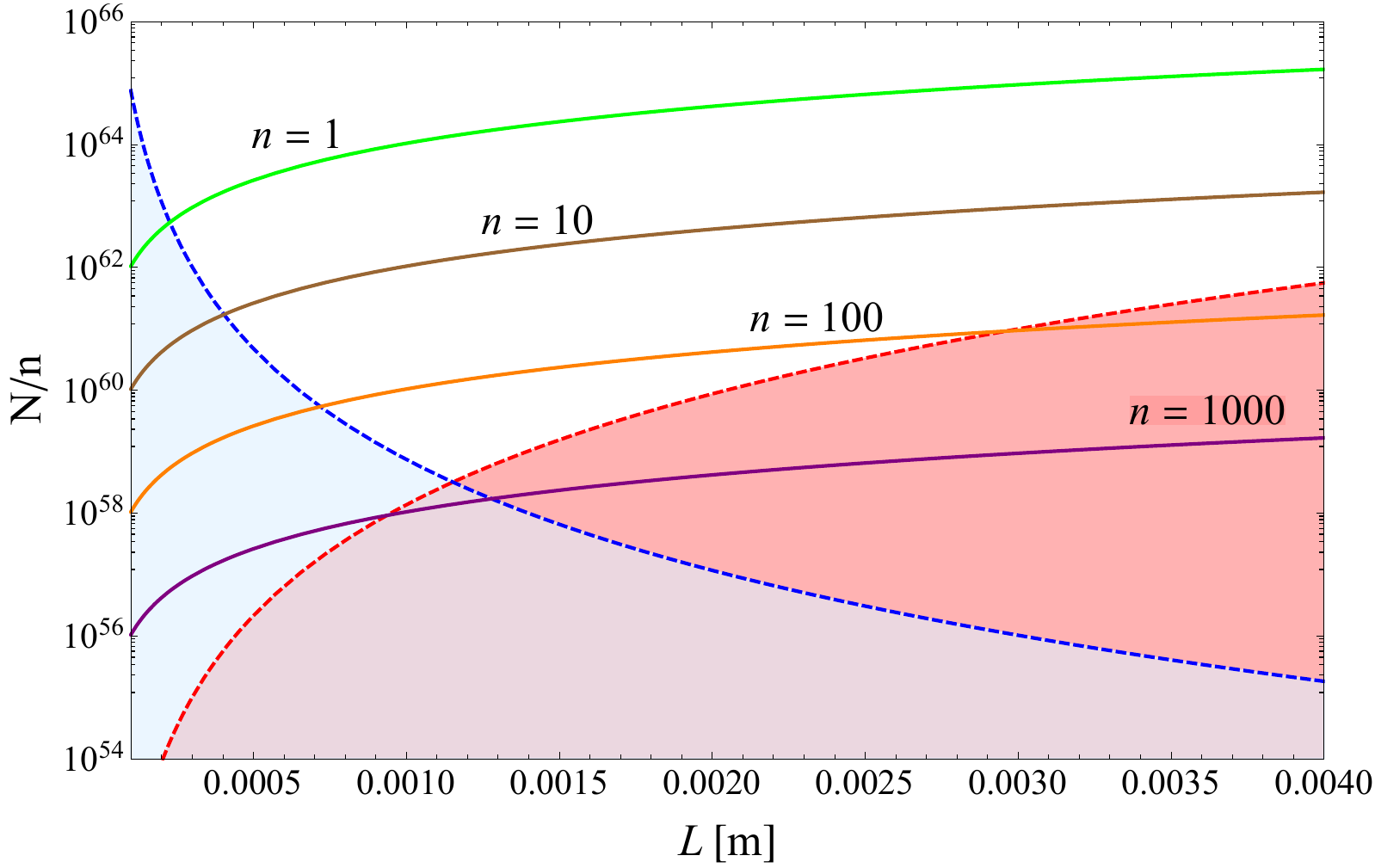}
  \caption{The dependence of the ratio $N/n$ on the AdS radius $L$ under various constraints at $g_s=0.1$. The dashed blue and red curves represent lower bounds that correspond to the observed value of the 4D cosmological constant $\rho_4\approx 10^{-48}$ GeV$^4$ \cite{Tanabashi2018} and the constraint on the extra dimensions coming from the supernova and neutron-star observations \cite{Hannestad2004}. The green, brown, orange, and purple curves refer to the 4D Planck scale corresponding to the number of escaping D3-branes $n=1$, $10$, $100$, and $1000$, respectively.}
  \label{fig:constk1}
\end{figure}

Then, we determine the 4D vacuum energy density $\rho_4$ in terms of the AdS radius $L$, the string coupling $g_s$, the 4D Planck scale $M_{\text{Pl}}$, and the ratio $n/N$ (for the $k=1$ case) as follows
\begin{eqnarray}
\rho_4=\left\{
		\begin{aligned}
			& \frac{32}{9}\sqrt{\frac{n}{N}}\frac{M_{\text{Pl}}}{g_sL^3} && \text{for} \quad k=1,\\
			& \frac{32}{3}\sqrt{\frac{1}{3}}\frac{M_{\text{Pl}}}{g_sL^3}  && \text{for} \quad k=0.
		\end{aligned}
		\right. \label{4DVED-corr}
\end{eqnarray}
The AdS radius $L$ is constrained by the observations of supernova and neutron star, which have been used to impose bounds on the size of extra dimensions. For $k=1$, we have a relation between the AdS radius $L$ and an equal radius $R$ for six extra dimensions which is used in Ref. \cite{Hannestad2004} as 
\begin{eqnarray}
 L=2\pi R\left(\frac{81N}{16\pi^3 n}\right)^{1/6},   
\end{eqnarray}
where an upper bound for $R$ was obtained as $R\lesssim4.41\times10^{-14}$ m coming from the constraints on the production of Kaluza-Klein gravitons in the supernova and neutron star. In Fig. \ref{fig:constk1}, we show constraints on the ratio $N/n$ and the AdS radius $L$ at $g_s=0.1$.
We find that with the string coupling around $0.1$, the ratio $N/n$ yielding the observed cosmological constant and satisfying the experimental constraints ranges from the order $10^{63}$ to $10^{58}$ corresponding to the AdS radius $L$ ranging from $2\times10^{-4}$ m to $\times10^{-3}$ m. This corresponds to the string length inverse $(\sqrt{\alpha'})^{-1}$ in the TeV scale, implying that quantum gravity effects can be observed in the low-energy regime \cite{Dimopoulos1998,Antoniadis1998}. In addition, we observe that the number of D-branes emitted from the stack must be below $\mathcal{O}(10^{3})$.

For $k=0$, we can find a relation between the AdS radius $L$ and the radius $R$ as
\begin{eqnarray}
L=2\pi R\left(\frac{27}{16\pi^3}\right)^{1/6}.    
\end{eqnarray}
This leads to an upper bound for the AdS radius $L$ from the supernova and neutron star limits as $L\lesssim1.71\times10^{-13}$ m. With $g_s\sim\mathcal{O}(0.1)$, we can find a minimum value for $\rho_4$ which is of order $10^{12}$ GeV$^4$, which is much larger than the experimental value. On the other hand, for the dS vacuum with $k=0$ to be consistent with the observations, there should be a fine-tuning between the positive contribution given by the second line in Eq. (\ref{4DVED-corr}) with a negative contribution which can come from the spontaneous symmetry breaking on the brane at the scale of the order of TeV. Interestingly, this energy scale is motivated by many unsolved problems in particle physics and cosmology.

\begin{figure}[ht]
  \includegraphics[scale=0.45]{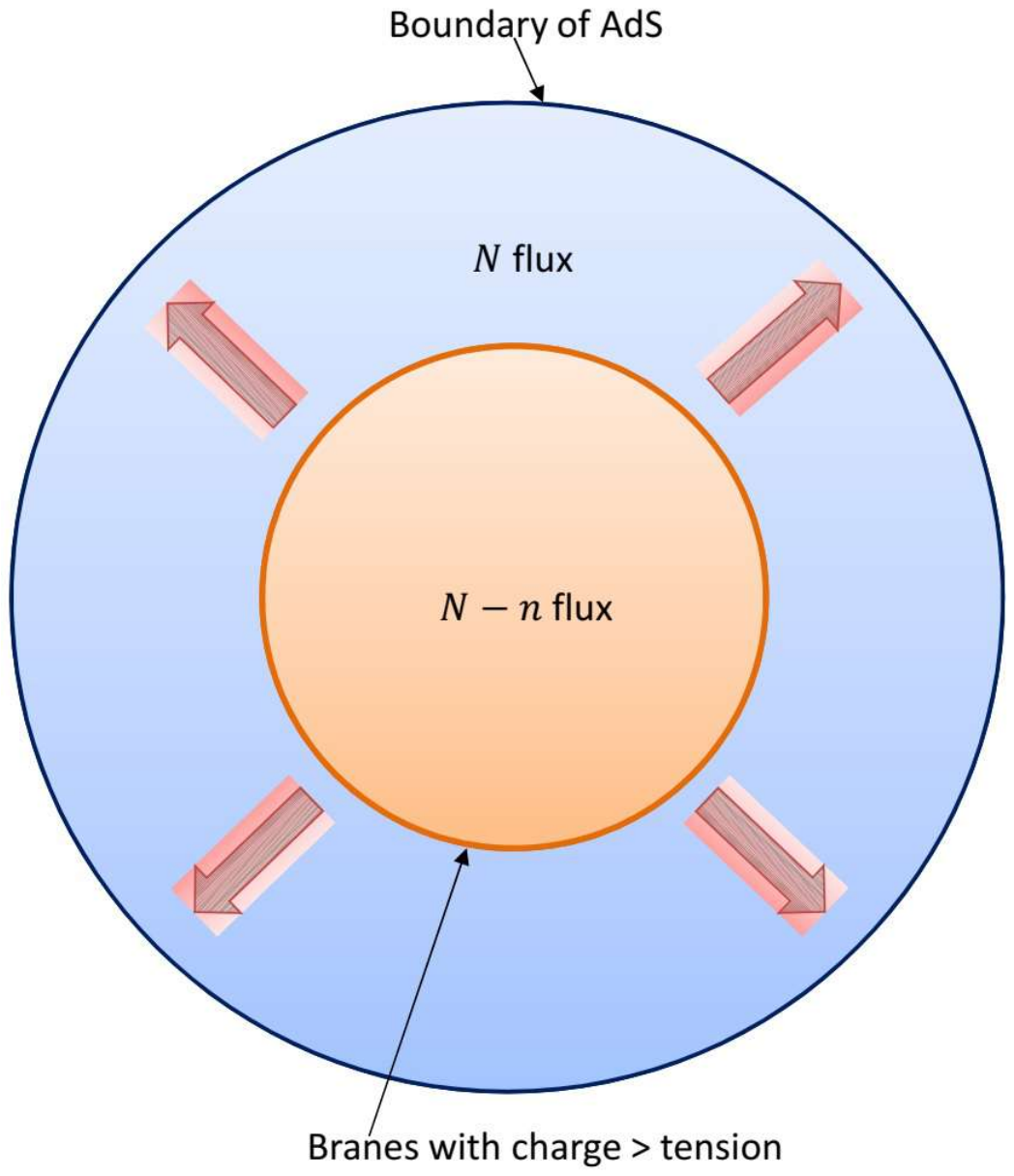}
  \caption{The orange circle refers to $n$ probe D3-branes, emitted from the stack of D3-branes placed at the conifold tip, which nucleate to mediate the decay of the metastable configuration. The light blue and light orange regions represent the metastable and stable configurations, respectively. The black circle refers to the AdS boundary.}
  \label{brane-nuc}
\end{figure}

Let us realize why the stringy corrections reduce the brane tension. In the present work, we consider a physical system that is determined by Eqs. (1), (2), and (3) that describe a U(1) charged black hole solution of IIB supergravity. 
\begin{figure}[ht]
  \includegraphics[scale=0.45]{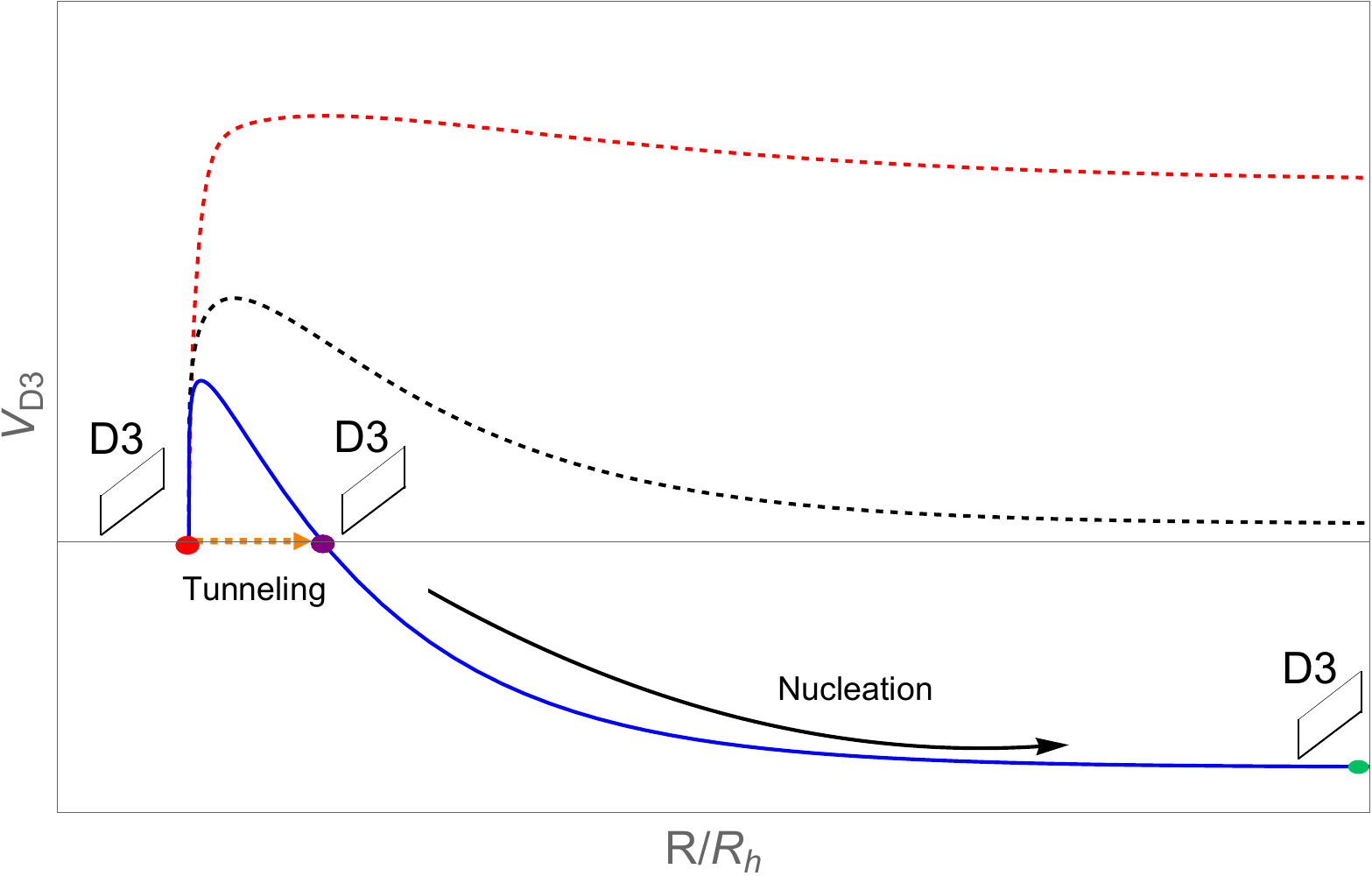}
  \caption{The effective potential $V_{D3}$ of a probe D3-brane in the rescaled radial coordinate $R/R_h$ with various values of $T/\mu$. The blue, dashed black, and dashed red curves correspond to the fact that the ratio $T/\mu$ is smaller than, equal to, and larger than the critical value $\approx0.2$, respectively.}
  \label{eff-pot}
\end{figure}
We assume that the physical system initially begins at a metastable configuration, which occurs when the chemical potential is sufficiently large and the black hole temperature is sufficiently low. In the metastable configuration, the effective potential for the probe D3-branes is given by the blue curve depicted in Fig. \ref{eff-pot} \cite{Henriksson2020}. Because a global minimum is located at $R\rightarrow\infty$, some D3-branes would be emitted from the stack of D3-branes placed at the conifold tip, tunnel through the potential barrier, and then nucleate to reach the AdS boundary as shown in Fig. \ref{brane-nuc}. 
To move towards the global minimum, the D3-branes emitted from the stack of D3-branes must lower their energy. For the D3-branes to nucleate, the Coulomb repulsive force due to their charge with respect to the five-form flux must overcome the attraction force exerted by gravity. This means that the tension of the nucleated D3-branes must be smaller than their charge, so-called super-extremal branes. Indeed, as we calculated above, the tension of the nucleated D3-branes is reduced via the stringy corrections. In this way, the presence of the super-extremal branes shows the form of non-zero stringy corrections in the metastable configuration. However, when the nucleated D3-branes reach the AdS boundary ($L/a\rightarrow0$), everything will be in the stable configuration which is represented by the light orange region given in Fig. \ref{brane-nuc}. In the stable configuration, the effective potential of the probe D3-branes is given by the red dashed curve shown in Fig. \ref{eff-pot}, where the global minimum is located inside the event horizon. As a result, when some branes are pulled out of the stack, they would tend to fall back into the initial position. This means that the extremal branes at the stack cannot decay by radiating the super-extremal branes, which means that stringy corrections to the brane tension should be zero in the stable configuration.

\section{\label{D5-nucle} Nucleated D5-brane}
In the previous section, we constructed the 4D dS vacuum on the worldvolume of the nucleated D3-brane where the matter fields (related to open strings), such as the SM fields and dark sector (new physics beyond the SM), do not experience the presence of extra dimensions. However, it is interesting to consider the possibility that some matter fields propagate in the compact extra dimensions because this situation can provide the solutions for cosmic inflation \cite{Nam2021}, the light neutrino masses \cite{Dienes1999}, and the dark matter \cite{Cembranos2003,Nam2024}. Therefore, in this section, we will construct the 4D dS vacuum on a nucleated D5-brane wrapping on a two-dimensional compact submanifold $M_2$ of the internal manifold $T^{1,1}$. As seen in the previous section, obtaining a tiny 4D cosmological constant requires a huge number of D3-branes placed at the conifold tip corresponding to the $L$ size of $T^{1,1}$ at the order of $10^{-4}$ m, for $k=1$. Whereas, the experimental constraints on the matter fields propagating in the extra dimensions would impose the size of $M_2$ that is (much) smaller than the radius $L$. Such a submanifold $M_2$ of $T^{1,1}$ is, in general, topologically trivial, hence the wrapped D5-brane would not be topologically stable. But, one can stabilize this configuration by using some dynamic mechanism such as the brane spinning on $T^{1,1}$ or Myers's mechanism \cite{Myers1999} where some D3-branes will polarize into the D5-brane wrapping on $M_2$ when a tachyon is present \cite{Polchinski2000}.

An essential point for the probe D5-brane, which can nucleate in the background determined by Eqs. (\ref{10D-back-metri}), (\ref{5D-back-metri}) and (\ref{C4-from-pot}), is the presence of a non-zero WZ term which produces a repulsive force. There is only the nonvanishing four-form potential $C_4$ in this background. This means that in order to have a non-zero WZ term, a U(1) gauge field corresponding to the field strength $F$ must be turned along the wrapping part $M_2$ of the D5-brane worldvolume, and some D3-branes charged under the four-form potential $C_4$ must dissolve in the D5-brane, leading to a bound state of D5-brane and D3-brane. (Note that the D3-branes dissolved in the D5-brane would also contribute to the field strength $F$. If the field strength $F$ is completely generated by the dissolved D3-brane charge, then the flux $\frac{1}{2\pi\alpha'}\int_{M_2}F$ would be proportional to the number of D3-branes which dilute in the D5-brane worldvolume \cite{Arean2006}.) 

As a result, the action of the D5-brane is given up to the corrections at order $\alpha'^2$ as follows
\begin{eqnarray}
S_{D5}&=&-T_5\int d^6\xi\sqrt{-\det(g_{\mu\nu}+F_{\mu\nu})}+S^{(\alpha'^2)}_{\text{cur.}}+S^{(\alpha'^2)}_F+T_5\int P[C_4]\wedge F\nonumber\\
&&\wedge\left[1+\frac{\pi^2\alpha'^2}{24}\left(\text{tr}\mathcal{R}^2_T-\text{tr}\mathcal{R}^2_N\right)\right],   
\end{eqnarray}
where $T_5=(2\pi)^{-5}/(g_s\alpha'^{3})$ is the tension of the D5-brane and $\xi^\mu$ (with $\mu=0,1,...,5$) are the D5-brane worldvolume coordinates. The curvature correction $S^{(\alpha'^2)}_{\text{cur.}}$ for the DBI action is similarly given as in Eq. (\ref{corr-D3-brane-act}). The correction $S^{(\alpha'^2)}_F$ which is related to the worldvolume field strength $F$ is given as
\begin{eqnarray}
S^{(\alpha'^2)}_F&=&-\frac{\pi^2\alpha'^2T_5}{12}\int d^6\xi\sqrt{|g|}\left[R_{\mu\nu}\left(\partial_\rho F^{\rho\mu}\partial_\sigma F^{\sigma\nu}-\partial_\rho{F_\sigma}^\nu\partial^\sigma F^{\rho\mu}\right)+\frac{1}{2}{R^{\mu\nu}}_{\rho\lambda}\partial_\mu F^{\delta\rho}\partial_\nu {F_\delta}^\lambda\right.\nonumber\\
&&\left.+\frac{1}{4}{R^{\mu\nu}}_{\mu\nu}\left(\partial_\mu F^{\mu\nu}\partial_\lambda{F_\nu}^\lambda+\partial_\mu{F_\nu}^\lambda\partial_\lambda F^{\nu\mu}\right)\right],
\end{eqnarray}
where $R_{\mu\nu}=g^{\rho\lambda}R_{\rho\mu\lambda\nu}$ and ${R^{\mu\nu}}_{\rho\lambda}=g^{\mu\mu'}g^{\nu\nu'}R_{\mu'\nu'\rho\lambda}$ with $R_{\mu\nu\rho\lambda}=R_{MNPQ}e^M_\mu e^N_\nu e^P_\rho e^Q_\lambda$.

Without loss of generality, we consider the D5-brane that wraps on a two-torus $T^2\subset T^{1,1}$. The embedding of the D5-brane in the 10D bulk is described by
\begin{eqnarray}
\mathcal{T}&=&\mathcal{T}(t), \ \  R=R(t), \ \ \Vec{x}\Big|_{\text{bulk}}=\Vec{x}\Big|_{\text{brane}},\ \ \Theta_{1,2}=\pi-\epsilon,\nonumber\\
\Phi_1&=&\phi_1+\phi_2,\ \ \Phi_2=-\phi_1+\phi_2,\ \ \Psi=2\phi_2,\label{D5-br-embe}
\end{eqnarray}
where $\epsilon$ with $|\epsilon|\ll1$ is a constant and the angles $\phi_{1,2}$ parameterize the two-torus $T^2$. The induced metric on the worldvolume of the wrapped D5-brane is
\begin{eqnarray}
ds^2_6&=&G_{MN}e^M_\mu e^N_\nu dx^\mu dx^\nu\nonumber\\
&=&e^{-5\chi(a)/3}\left(-\gamma_{tt}dt^2+a^2d\Omega^2_{3,k}\right)+e^{\chi(a)+\eta(a)}\left(R^2_1d\phi^2_1+R^2_2d\phi^2_2\right),\label{D6-br-metri}
\end{eqnarray}
where the radii $R_{1,2}$ is given by
\begin{eqnarray}
R_1&=&\frac{L}{\sqrt{3}}|\sin\epsilon|,\\    
R_2&=&\frac{2L}{3}\sqrt{\frac{3}{4}\sin^2\epsilon+e^{-5\eta(a)}(1-\cos\epsilon)^2},
\end{eqnarray}
and we have considered $k=1$, which can yield a tiny cosmological constant as indicated above. For the small $\epsilon$ angle, we have $R_1\simeq R_2$. The field strength $F$ of the worldvolume U(1) gauge field along the two-torus $T^2$ is given by 
\begin{eqnarray}
F=q d\phi_1\wedge d\phi_2,
\end{eqnarray}
where $q$ is a positive constant.

With the induced metric of the D5-brane worldvolume given by E. (\ref{D6-br-metri}), we obtain 
\begin{eqnarray}
\left(R_T\right)_{\mu\nu\rho\lambda}\left(R_T\right)^{\mu\nu\rho\lambda}-\left(R_T\right)_{\mu\nu}\left(R_T\right)^{\mu\nu}&=&-24\left[\frac{k}{2a^4}+\left(\frac{1}{a^2}+\frac{24\eta'(a)}{a}\right)\left(\frac{\dot{a}}{a}\right)^2+\frac{1}{a^2}\frac{\ddot{a}}{a}\right.\nonumber\\
&&\left.+\mathcal{O}\left(\frac{\dot{a}^4}{a^4},\frac{\ddot{a}^2}{a^2},\frac{\dot{a}^2\ddot{a}}{a^3}\right)\right].
\end{eqnarray}
The remaining curvature corrections for the D5-brane action are expanded as follows
\begin{eqnarray}
\left(R_N\right)_{\mu\nu ab}\left(R_N\right)^{\mu\nu ab}&=&\frac{16}{L^4}\left[1-\frac{R^2_1}{L^2}+\mathcal{O}\left(\frac{R^4_1}{L^4}\right)\right]-\frac{16Q^2}{5L^8}\left[1-\frac{47R^2_1}{32L^2}+\mathcal{O}\left(\frac{R^4_1}{L^4}\right)\right]\left(\frac{L}{a}\right)^5\nonumber\\
&&+\mathcal{O}\left(\frac{L^6}{a^6}\right),\\
\bar{R}_{ab}\bar{R}^{ab}&=&\frac{8}{R^4_1}\left[1-\frac{2R^2_1}{L^2}+\mathcal{O}\left(\frac{R^4_1}{L^4}\right)\right]-\frac{Q^2}{5L^4R^4_1}\left[1+\frac{3R^2_1}{L^2}+\mathcal{O}\left(\frac{R^4_1}{L^4}\right)\right]\left(\frac{L}{a}\right)^5\nonumber\\
&&+\mathcal{O}\left(\frac{L^6}{a^6}\right),
\end{eqnarray}
whose detailed computations are given in Appendix B. At the late time ($L/a\ll1$), we have the following hierarchy
\begin{eqnarray}
\frac{\dot{a}}{a},\left(\frac{\ddot{a}}{a}\right)^{1/2},\frac{1}{a}\ll\frac{1}{L}\ll\frac{1}{R_1}    
\end{eqnarray}
This means that the contribution of $\bar{R}_{ab}\bar{R}^{ab}$ is much larger than that of $\left(R_N\right)_{\mu\nu ab}\left(R_N\right)^{\mu\nu ab}$ and $\left(R_T\right)_{\mu\nu\rho\lambda}\left(R_T\right)^{\mu\nu\rho\lambda}-\left(R_T\right)_{\mu\nu}\left(R_T\right)^{\mu\nu}$. In addition, the correction $S^{(\alpha'^2)}_F$ vanishes because the field strength tensor $F$ is constant. Hence, the stringy corrections to the D5-brane action wrapped on the two-torus $T^2$ should come from the term $\bar{R}_{ab}\bar{R}^{ab}$. 

Finally, we find the D5-brane action including the stringy corrections as follows
\begin{eqnarray}
S_{5D}&\simeq&-4\pi^2T_5V_3\int dt\sqrt{\gamma_{tt}}\left\{a^3\left(\sqrt{q^2+R^4_1}-\frac{\pi^2}{3}\frac{\alpha'^2}{R^2_1}\right)-\frac{q}{L}\left[a^4-\frac{Q^2L^2}{15a^2}-\left(R^4_h-\frac{Q^2L^2}{15R^2_h}\right)\right]
\dot{\mathcal{T}}\right\}\nonumber\\
&\equiv&\int dt\mathcal{L}_{D5}.    
\end{eqnarray}
The equation of motion for the D5-brane leads to the Friedmann equation describing its expansion at the late time as follows
\begin{eqnarray}
\left(\frac{\dot{a}}{a}\right)^2+\frac{1}{a^2}\simeq\frac{1}{L^2}\left[\frac{q^2}{\left(\sqrt{q^2+R^4_1}-\frac{\pi^2}{3}\frac{\alpha'^2}{R^2_1}\right)^2}-1\right]+\frac{L^2}{2}\left(\frac{Q^2}{15L^2R^2_h}-\frac{R^4_h}{L^4}-2g_2\right)\frac{1}{a^4}.
\end{eqnarray}
In order to have a positive 4D cosmological constant, the $q$ parameter needs to satisfy the following condition
\begin{eqnarray}
q>\frac{3R^2_1}{2\pi^2}\frac{R^4_1}{\alpha'^2}\left[1-\left(\frac{\pi^2\alpha'^2}{3R^4_1}\right)^2\right]\simeq\frac{3R^4_1}{2\pi^2\alpha'^2}R^2_1.   
\end{eqnarray}
This condition means that the parameter $\sqrt{q}$ related to the field strength of the worldvolume U(1) gauge field needs to be much larger than the size of the two-torus $T^2$ upon which the D5-brane wraps. The 4D vacuum energy density reads
\begin{eqnarray}
\rho_4\simeq\frac{2M^2_{\text{Pl}}}{qL^2}\left(\frac{\pi^2\alpha'^2}{3R^2_1}-\frac{R^4_1}{2q}\right). 
\end{eqnarray}
With the observed value $\rho_4\sim10^{-48}$ GeV$^4$, we can estimate the order of $q$ as follows
\begin{eqnarray}
\frac{q}{\text{GeV}^{-2}}\sim10^{48}\left(\frac{\text{GeV}^{-1}}{R_1}\right)^2.   
\end{eqnarray}
If the SM fields propagate in the two-torus $T^2$, then a lower bound for the inverse radius $(R_1)^{-1}$ is of order TeV. This leads to a lower bound for $q$ as $q\gtrsim\mathcal{O}(10^{54})$ GeV$^{-2}$. On the contrary, if the SM fields live on the D3-branes diluting in the D5-brane and the dark sector can only propagate in $T^2$, then a lower bound for $(R_1)^{-1}$ is possibly much lower than the TeV scale. For instance, with $(R_1)^{-1}\sim10^{-10}$ GeV, the value of $q$ which needs to lead to the observed cosmological constant is the order of $10^{28}$ GeV$^{-2}$.

\section{\label{sec:conclu} Conclusion}
We construct 4D dS vacua consistent with the observations on the worldvolume of D-branes nucleating in a 10D stringy background, which is holographically dual to the gauge theory with a finite density and nonzero temperature. In the situation that the chemical potential is high enough and the black hole temperature is low enough, the effective potential for a probe D-brane exhibits a global minimum that is located at $R\rightarrow\infty$ with $R$ being the radial coordinate \cite{Herzog2010,Henriksson2020}. Hence, the system would emit some D-branes to reach the global minimum. First, we study the nucleation of a probe D3-brane. The computation indicates that the cosmological constant vanishes on the D3-brane worldvolume at the leading order of the string length. However, we show that when stringy corrections are included, they would lead to a small and nonvanishing cosmological constant without being fine-tuned for the case of the spherical D3-brane.

In order to incorporate the matter fields that propagate in the compact extra dimensions, we study the nucleation of a probe D5-brane wrapping on a two-torus of the internal manifold $T^{1,1}$. The repulsive force for the expansion of the D5-brane comes from the Chern-Simons coupling between the field strength two-form of the worldvolume U(1) gauge field and the four-form gauge potential of the 10D background. We show that stringy corrections and a sufficiently large value of the field strength two-form are essential to produce a dS vacuum that is consistent with the observations.

\section*{Acknowledgments}
We appreciate JINR (Dubna) for the hospitality where part of this work was done.

\section*{\label{D3-append} Appendix A: Computing the scalar curvatures for the D3-brane}

In order to compute $\left(R_N\right)_{\mu\nu ab}\left(R_N\right)^{\mu\nu ab}$, we express $\left(R_N\right)_{\mu\nu ab}$ in the following form 
\begin{eqnarray}
\left(R_N\right)_{\mu\nu ab}=\widetilde{R}_{MN,\mu\nu}E^M_aE^N_b,
\end{eqnarray}
where $\widetilde{R}_{MN,\mu\nu}$ is defined by
\begin{eqnarray}
\widetilde{R}_{MN,\mu\nu}=-R_{MNPQ}e^P_\mu e^Q_\nu+g^{\rho\lambda}\left(\Omega^P_{\mu\rho}\Omega^Q_{\nu\lambda}-\Omega^Q_{\mu\rho}\Omega^P_{\nu\lambda}\right)G_{PM}G_{QN}.\label{wid-RMNmn}
\end{eqnarray}
With the embedding of the D3-brane given by Eq. (\ref{D3-br-embe}), the non-zero components of $\Omega^M_{\mu\nu}$ are
\begin{eqnarray}
\Omega^R_{tt}&=&\ddot{a}-\left(\Gamma_T\right)^t_{tt}\dot{a}+\Gamma^R_{\mathcal{T}\mathcal{T}}\dot{\mathcal{T}}^2+\Gamma^R_{RR}\dot{a}^2\label{Omg-D3}\\
&=&\frac{1}{2}\left[g'(a)-g(a)\left(w'(a)+\frac{5}{3}\chi'(a)\right)\right]+\mathcal{O}\left(\frac{\dot{a}^2}{a^2},\frac{\ddot{a}}{a}\right)\nonumber\\
\Omega^R_{x^ix^i}&=&-\left(\Gamma_T\right)^t_{x^ix^i}\dot{a}+\Gamma^R_{x^ix^i}\nonumber\\
&=&-ag(a)\left[1-\frac{5}{6}a\chi'(a)\right]f_{x^i}+\mathcal{O}\left(\frac{\dot{a}^2}{a^2},\frac{\ddot{a}}{a}\right),\nonumber
\end{eqnarray}
where the functions $f_{x^i}$ are given by
\begin{eqnarray}
f_{x^i}=\left\{
		\begin{aligned}
			& 1 && \text{for} \quad x^i=\psi,\\
			&  \sin^2\psi && \text{for} \quad x^i=\theta,\\
                &  \sin^2\psi\sin^2\theta && \text{for} \quad x^i=\phi,
		\end{aligned}
		\right. 
\end{eqnarray}
in the case of $k=1$ and $f_{x^i}=1$ in the case of $k=0$. Because $\Omega^M_{\mu\nu}$ is non-zero for $\mu=\nu$ as found in Eq. (\ref{Omg-D3}) and the metric $g_{\mu\nu}$ is diagonal, the second term in Eq. (\ref{wid-RMNmn}) is vanishing, hence we obtain $\widetilde{R}_{MN,\mu\nu}=-R_{MNPQ}e^P_\mu e^Q_\nu$. Finally, we find
\begin{eqnarray}
\left(R_N\right)_{\mu\nu ab}\left(R_N\right)^{\mu\nu ab}&=&g^{\mu\rho}g^{\nu\lambda}\delta^{ac}\delta^{bd}\widetilde{R}_{MN,\mu\nu}E^M_aE^N_b\widetilde{R}_{PQ,\rho\lambda}E^P_cE^Q_d\nonumber\\
&=&g^{\mu\rho}g^{\nu\lambda}\bar{G}^{MP}\bar{G}^{NQ}R_{MNM'N'}e^{M'}_\mu e^{N'}_\nu R_{PQP'Q'}e^{P'}_\rho e^{Q'}_\lambda,\nonumber\\
&=&0,
\end{eqnarray}
where $\bar{G}^{MP}$ is defined as $\bar{G}^{MP}\equiv E^M_aE^{P}_c\delta^{ac}=G^{MP}-e^M_\mu e^{P}_\nu g^{\mu\nu}$ whose non-zero component is
\begin{eqnarray}
\bar{G}^{RR}&\simeq&g(a)e^{\frac{5\chi(a)}{3}}.\label{barG-D3}
\end{eqnarray}
Note that the last line has been obtained by using a fact that $\bar{G}^{MP}$ is non-zero only for the indices $M, P$ corresponding to the radial coordinate of the AdS space, and the Riemann tensor is antisymmetric as $R_{MNPQ}=-R_{MNQP}$. 

In analogy, we can compute $\bar{R}_{ab}\bar{R}^{ab}$ as follows
\begin{eqnarray}
\bar{R}_{ab}\bar{R}^{ab}&=&\bar{G}^{MP}\bar{G}^{NQ}\hat{R}_{MN}\hat{R}_{PQ},
\label{Ads-D3-Rba}       
\end{eqnarray}
where $\hat{R}_{MN}$ is defined by
\begin{eqnarray}
\hat{R}_{MN}=g^{\mu\nu}R_{PMQN}e^P_\mu e^Q_\nu+g^{\mu\nu}g^{\rho\lambda}\left(\Omega^P_{\mu\rho}\Omega^Q_{\nu\lambda}-\Omega^P_{\mu\nu}\Omega^Q_{\rho\lambda}\right)G_{PM}G_{QN}.\label{def-Rhat}     
\end{eqnarray}
Then, we find
\begin{eqnarray}
\bar{R}_{ab}\bar{R}^{ab}&=&\frac{e^{\frac{10 \chi (a)}{3}}}{144 a^4 g(a)^2}\left\{2 a g(a) \left[5 a \chi '(a)-6\right]\left[3 g'(a)-3 a g w'(a)+5 a g \chi '(a)\right]-2 g(a)^2 \left[6-5 a \chi '(a)\right]^2\right.\nonumber\\
&&\left.+ag(a)\left[g'(a) \left(9 a w'(a)+25 a \chi '(a)-18\right)+a\left(6 w''(a)+40 \chi ''(a)-5 \chi '(a) w'(a)-3 w'(a)^2\right)\right.\right.\nonumber\\
&&\left.\left.-6 a g''(a)\right]\right\}.
\end{eqnarray}

\section*{Appendix B: Computing the scalar curvatures for the D5-brane}
With the embedding of the D5-brane given by Eq. (\ref{D5-br-embe}) and the 10D bulk metric given in Eq. (\ref{10D-back-metri}), we can compute the non-zero components of $\Omega^M_{\mu\nu}$. The non-zero components that are related to the non-compact part of the D5-brane worldvolume are given in Eq. (\ref{Omg-D3}). Whereas, the non-zero components that are related to the compact part and the mixing between the non-compact and compact parts are
\begin{eqnarray}
\Omega^{R}_{\phi_1\phi_1}&=&-\frac{L^2}{6}g(a)e^{\eta(a)+\frac{8\chi(a)}{3}}\left[\eta'(a)+\chi'(a)\right]\sin^2\epsilon,\nonumber\\
\Omega^{R}_{\phi_2\phi_2}&=&\frac{L^2}{18}g(a)e^{\frac{8\chi (a)}{3}-4\eta (a)}\left[4 (\cos\epsilon-1)^2\left(4\eta'(a)-\chi'(a)\right)\right.\nonumber\\
&&\left.-3 e^{5\eta(a)}\left(\eta'(a)+\chi'(a)\right)\sin^2\epsilon\right],\nonumber\\
\Omega^{\Theta_1}_{\phi_1\phi_1}&=&\frac{1}{2}\sin2\epsilon,\nonumber\\
\Omega^{\Theta_1}_{\phi_1\phi_2}&=&\Omega^{\Theta_1}_{\phi_2\phi_1}=\frac{1}{2}\sin2\epsilon-\frac{2}{3}e^{-5\eta (a)}(\cos\epsilon-1)\sin\epsilon,\\
\Omega^{\Theta_1}_{\phi_2\phi_2}&=&\frac{1}{2}\sin2\epsilon-\frac{4}{3}e^{-5\eta(a)}(\cos\epsilon-1)\sin\epsilon,\nonumber\\
\Omega^{\Theta_2}_{\phi_1\phi_1}&=&\frac{1}{2}\sin2\epsilon,\nonumber\\
\Omega^{\Theta_2}_{\phi_1\phi_2}&=&\Omega^{\Theta_2}_{\phi_2\phi_1}=-\frac{1}{2}\sin2\epsilon+\frac{2}{3}e^{-5\eta (a)}(\cos\epsilon-1)\sin\epsilon,\nonumber\\
\Omega^{\Theta_2}_{\phi_2\phi_2}&=&\frac{1}{2}\sin2\epsilon-\frac{4}{3}e^{-5\eta(a)}(\cos\epsilon-1)\sin\epsilon.\nonumber
\end{eqnarray}
The non-zero components of $\bar{G}^{MP}=G^{MP}-e^M_\mu e^{P}_\nu g^{\mu\nu}$ corresponding to the embedding of the D5-brane read
\begin{eqnarray}
\bar{G}^{RR}&\simeq&g(a)e^{\frac{5\chi(a)}{3}},\nonumber\\
\bar{G}^{\Theta_1\Theta_1}&=&\bar{G}^{\Theta_2\Theta_2}=\frac{6}{L^2}e^{-\eta (a)-\chi (a)},\nonumber\\
\bar{G}^{\Phi_1\Phi_1}&=&\bar{G}^{\Phi_2\Phi_2}=\bar{G}^{\Phi_1\Phi_2}=\bar{G}^{\Phi_2\Phi_1}=\frac{6e^{-\eta(a)-\chi(a)}\sec^2\left(\frac{\epsilon }{2}\right)}{L^2\left[3e^{5\eta(a)} (\cos\epsilon+1)-4\cos\epsilon+4\right]},\\
\bar{G}^{\Phi_1\Psi}&=&\bar{G}^{\Phi_2\Psi}=\frac{3e^{-\eta (a)-\chi (a)}}{2 L^2}\left[4 \cot\epsilon\csc\epsilon-\frac{3}{e^{-5\eta (a)} (\cos\epsilon-1)^2+\frac{3\sin^2\epsilon}{4}}\right],\nonumber\\
\bar{G}^{\Psi\Psi}&=&\frac{3e^{-\eta (a)-\chi (a)}\left[\left(3e^{5\eta(a)}-4\right)\sin\epsilon+4\tan\left(\frac{\epsilon }{2}\right)\right]^2}{L^2\left[3e^{5\eta(a)}\sin^2\epsilon+4(\cos\epsilon-1)^2\right]},\nonumber
\end{eqnarray}
where $\sec z=1/\cos z$ and $\csc z=1/\sin z$. Then, we can compute $\left(R_N\right)_{\mu\nu ab}\left(R_N\right)^{\mu\nu ab}$ as follows
\begin{eqnarray}
\left(R_N\right)_{\mu\nu ab}\left(R_N\right)^{\mu\nu ab}&=&g^{\mu\rho}g^{\nu\lambda}\bar{G}^{MP}\bar{G}^{NQ}\widetilde{R}_{MN,\mu\nu}\widetilde{R}_{PQ,\rho\lambda},\nonumber\\
&=&\frac{25g(a)e^{\frac{2\chi (a)}{3}-6\eta(a)}\eta'(a)^2}{L^2\left[3e^{5\eta(a)}\sin^2\epsilon+4(\cos\epsilon-1)^2\right]^3} 
\left[9e^{10\eta (a)}\left(39-47\cos\epsilon+32\cos2\epsilon-9\cos3\epsilon\right.\right.\nonumber\\
&&\left.\left.+\cos4\epsilon\right)-48e^{5\eta (a)}\sin^2\left(\frac{\epsilon }{2}\right)\left(\cos2\epsilon-7\cos\epsilon\right)+128\sin ^4\left(\frac{\epsilon}{2}\right)\right]\sin^4\left(\frac{\epsilon }{2}\right)\nonumber\\
&&+\frac{12e^{-15\eta(a)-2\chi (a)}}{L^4(\cos\epsilon-7)^2\left[3e^{5\eta (a)}+4\tan^2\left(\frac{\epsilon }{2}\right)\right]}
\left[2e^{\chi(a)}\sec^2\left(\frac{\epsilon }{2}\right)\left(\left(e^{5\eta(a)}-1\right)\cos\epsilon-1\right)\right.\nonumber\\
&&\left.\times\left(\left(3e^{5\eta (a)}-2\right)\cos\epsilon+2\right)+e^{4\eta (a)}(\cos\epsilon-7)\right]^2.
\end{eqnarray}

The scalar curvature $\bar{R}_{ab}\bar{R}^{ab}$ is computed as follows 
\begin{eqnarray}
\bar{R}_{ab}\bar{R}^{ab}&=&\bar{G}^{MP}\bar{G}^{NQ}\hat{R}_{MN}\hat{R}_{PQ}\nonumber\\
&=&-\frac{90g(a)e^{\frac{2\chi(a)}{3}-\eta(a)}\eta'(a)}{L^2\left[3e^{5\eta(a)}(\cos\epsilon+1)-4\cos\epsilon+4\right]}\sin\epsilon+\frac{e^{-2(6\eta(a)+\chi (a))}}{2L^4\left[3e^{5\eta(a)}(\cos\epsilon+1)-4\cos\epsilon+4\right]^2}\nonumber\\
&&\times\left[18e^{10\eta (a)}\left(121-20\cos\epsilon+99\cos2\epsilon\right)-96e^{5\eta(a)}\sin^2\left(\frac{\epsilon }{2}\right)(13\cos \epsilon-3)\right.\nonumber\\
&&\left.+81 e^{20\eta(a)}(\cos4\epsilon+3)\csc^4\left(\frac{\epsilon }{2}\right)+54e^{15\eta (a)}(17\cos\epsilon-2\cos2\epsilon+15\cos3 \epsilon+2)\csc^2\left(\frac{\epsilon }{2}\right)\right.\nonumber\\
&&\left.+256\sin^4\left(\frac{\epsilon }{2}\right)\right].
\end{eqnarray}

\end{document}